\documentclass[aps,twocolumn,pre]{revtex4}
\usepackage{graphicx}
\usepackage{amsfonts}
\usepackage{amssymb}
\usepackage{amsmath}

\begin{document}

\title{\bf Symmetry Breaking in Symmetric and Asymmetric Double-Well Potentials}

\author{ 
G.\ Theocharis$^{1}$, 
P.G.\ Kevrekidis$^{2}$,  
D.J.\ Frantzeskakis$^{1}$, and 
P.\ Schmelcher$^{3,4}$ 
}
\affiliation{
$^{1}$ Department of Physics, University of Athens, Panepistimiopolis,Zografos, Athens 157 84, Greece \\
$^{2}$ Department of Mathematics and Statistics,University of Massachusetts, Amherst MA 01003-4515, USA \\
$^{3}$ Theoretische Chemie, Physikalisch-Chemisches Institut, INF 229, Universit\"at Heidelberg, 69120 Heidelberg, Germany \\
$^{4}$ Physikalisches Institut, Philosophenweg 12, Universit\"at Heidelberg, 69120 Heidelberg, Germany 
}

\begin{abstract}

Motivated by recent experimental studies of matter-waves and optical beams 
in double well potentials, we study the solutions of 
the nonlinear Schr\"{o}dinger equation in such a context.  
Using a Galerkin-type approach, 
we obtain a detailed handle on the nonlinear solution branches of the problem, 
starting from the corresponding linear ones and predict the relevant bifurcations
of solutions for both attractive and repulsive nonlinearities. The results 
illustrate the nontrivial differences that arise between the 
steady states/bifurcations emerging in symmetric and asymmetric double wells.

\end{abstract}

\maketitle 



{\it Introduction}. 
It is well known that the nonlinear Schr\"{o}dinger (NLS) equation is a 
fundamental 
model describing the evolution of a complex field envelope in nonlinear dispersive media \cite{NLS}. 
As such, it plays a key role in many different contexts, ranging from nonlinear and atom optics to 
plasma physics, fluid dynamics, and even biophysical models \cite{trubatch}.
The interest in the NLS equation has dramatically increased during the last 
few years, as it also describes the mean-field dynamics of Bose-Einstein condensates (BECs) \cite{gbec}. 
In this context, the NLS is also known as the 
Gross-Pitaevskii (GP) equation, and
typically incorporates external potentials that are used 
for the BEC confinement. Such potentials may be, e.g., harmonic (usually implemented 
by external magnetic fields) or periodic (implemented by the 
interference of laser beams), 
so-called optical lattices \cite{reviewsbec}. 
Importantly, NLS models with 
similar external potentials appear also in the 
context of optics, where they respectively describe the evolution of an optical beam in a 
graded-index waveguide or in periodic waveguide arrays \cite{kivshar,reviewsopt}. 

Another type of external potential, which has mainly been studied 
theoretically in the BEC context \cite{smerzi,kiv2,mahmud,bam,Bergeman_2mode,infeld} 
is the double well potential. 
Moreover, it has been demonstrated experimentally that a BEC 
either tunnels and performs Josephson oscillations between the wells, or is  
subject to macroscopic quantum self-trapping \cite{markus1}.
On the other hand, in the context of optics, a double well potential can be created 
by a two-hump self-guided laser beam in Kerr media \cite{HaeltermannPRL02}.
A different alternative was offered in \cite{zhigang}, wherein the first stages 
of the evolution of an optical beam, initially focused between the wells of a photorefractive 
crystal, were monitored. 
 
One of the particularly interesting features of either matter-waves or optical beams in 
double well potentials is the {\it spontaneous symmetry breaking}, i.e., the localization
of the respective order parameter in one of the wells of the potential. 
Symmetry breaking solutions of the NLS model have first been predicted 
in the context of molecular states \cite{ms} and, apart from the physical contexts 
of BECs \cite{smerzi,kiv2,mahmud,bam,Bergeman_2mode,infeld} (see also \cite{jackson}) 
and optics \cite{HaeltermannPRL02,zhigang} mentioned above, they 
have also been studied from a mathematical point of view in Refs. 
\cite{mathsy,mathsy1}. 

These works underscore the relevance and timeliness of a better 
understanding of the dynamics of nonlinear waves in double well potentials.  
In view of that, in the present work we offer a systematic methodology, 
based on a two-mode expansion, of how to tackle problems in double wells, 
as regards their stationary states and the bifurcations (and ensuing instabilities) 
that arise in them. This way, considering both cases of attractive and repulsive nonlinearities, 
we illustrate the ways in which a symmetric double well potential is different 
from an asymmetric one. In particular, we demonstrate that, 
contrary to the case of symmetric potentials where symmetry 
breaking follows a pitchfork bifurcation, 
in asymmetric double wells the bifurcation is of the saddle-node type.

The paper is structured as follows: in Section II, 
we present the model and set the analytical framework.
In Section III, we illustrate the value of the method by 
highlighting the significant differences of symmetric and asymmetric double wells. Finally,
in Section IV, we summarize our findings and discuss future directions.

{\it Model and Analytical Approach}.
In a quasi-1D setting, the evolution of the mean-field wavefunction of a BEC 
\cite{reviewsbec} (or the envelope of an optical beam \cite{kivshar})  
is described by the following normalized NLS (GP) equation, 
\begin{eqnarray}
i u_t=-\frac{1}{2} u_{xx} + s |u|^2 u + V(x) u - \mu u.
\label{beq1}
\end{eqnarray}
In the BEC (optics) context, $\mu$ denotes the chemical potential (propagation constant) and 
$s=\pm 1$ for attractive or repulsive interatomic interactions (focusing or defocusing Kerr nonlinearity) 
respectively; below, for simplicity, we will adopt the terms attractive and repulsive nonlinearity 
for $s=\pm 1$ respectively. Finally, in Eq. (\ref{beq1}), $V(x)$ is the double well potential, 
which is assumed to be composed by a parabolic trap 
(of strength $\Omega$) and a ${\rm sech}^2$-shaped barrier (of strength $V_0$, width $w$ and location $x_0$); 
in particular, $V(x)$ is of the form:
\begin{eqnarray}
V(x)=\frac{1}{2} \Omega^2 x^2 + V_0 {\rm sech}^2 \left(\frac{x-x_{0}}{w}\right),  
\label{beq2}
\end{eqnarray}
with the choice $x_{0}=0$ ($x_{0} \neq 0$) corresponding to a symmetric (asymmetric) double well. 
Note that such a double well can be implemented in BEC experiments 
upon, e.g., combining a magnetic trap with a sharply focused, blue-detuned laser beam \cite{expd}.  
Similar double wells can also be implemented e.g., in optical systems.

The spectrum of the underlying linear Schr{\"o}dinger equation ($s=0$) 
consists 
of a ground state, $u_{0}(x)$, and excited states, $u_{l}(x)$ ($l \ge 1$). In the 
nonlinear problem, using a Galerkin-type approach, we expand $u(x,t)$ as,
\begin{eqnarray}
u(x,t)=c_0(t) u_0(x) + c_1(t) u_1(x)+\cdots,
\label{beq3}
\end{eqnarray}
and truncate the expansion, keeping solely the first two modes; 
here $c_{0,1}(t)$ are unknown time-dependent complex prefactors. It is worth noticing that 
such an approximation (involving the truncation of higher order modes and 
the spatio-temporal factorization of the wavefunction), 
is expected to be quite useful for a weakly nonlinear analysis. In fact, as will be seen below, 
we will be able to identify the nonlinear states that stem from the linear ones, as well as their 
bifurcations.

Substituting Eq. (\ref{beq3}) into Eq. (\ref{beq1}), 
and projecting the result to the corresponding eigenmodes, we 
obtain the following ordinary differential equations (ODEs):
\begin{eqnarray}
i \dot{c}_0 &=& (\omega_0-\mu) c_0 - s A_0 |c_0|^2 c_0 - s B (2 |c_1|^2 c_0
+ c_1^2 \bar{c}_0), 
\nonumber 
\\
&-& s \Gamma_1 |c_1|^2 c_1  - s \Gamma_0  (2 |c_0|^2 c_1 + c_0^2 \bar{c}_1)
\label{beq4}
\\
i \dot{c}_1 &=& (\omega_1-\mu) c_1 - s A_1 |c_1|^2 c_1 - s B (2 |c_0|^2 c_1
+ c_0^2 \bar{c}_1), 
\nonumber 
\\
&-& s \Gamma_0 |c_0|^2 c_0  - s \Gamma_1  (2 |c_1|^2 c_0 + c_1^2 \bar{c}_0).
\label{beq5}
\end{eqnarray}
In Eqs. (\ref{beq4})-(\ref{beq5}), dots denote time derivatives, 
overbars denote complex conjugates, $\omega_{0,1}$ are the eigenvalues 
corresponding to the eigenstates $u_{0,1}$, while $A_0=\int u_0^4 dx$, $A_1=\int u_1^4 dx$, $B=\int u_0^2 u_1^2 dx$,
$\Gamma_0=\int u_0 u_1^3 dx$ and $\Gamma_1=\int u_1 u_0^3 dx$ are constants.
Recall that $u_{0}$ and $u_1$ are real (due to the Hermitian nature of
the underlying linear Schr{\"o}dinger problem) and are also orthonormal.
Notice also that in the symmetric case ($x_{0}=0$), due to the parity of the eigenfunctions, $\Gamma_0=\Gamma_1=0$.

We now use amplitude-phase (action-angle) variables, $c_j=\rho_j e^{i \phi_j}$, $j=0,1$
($\rho_j$ and $\phi_j$ are assumed to be real), to 
derive from the ODEs (\ref{beq4})-(\ref{beq5}) a set of four equations. 
Introducing the function $\varphi \equiv \phi_1-\phi_0$, we find that the equations for 
$\rho_0$ and $\phi_0$ are,
\begin{eqnarray}
\dot{\rho}_0 &=& s [(\Gamma_0 \rho_0^2
+ \Gamma_1 \rho_1^2) \sin(\varphi) - \rho_1^2 \rho_0 \sin(2 \varphi)], 
\label{beq6}
\\
\dot{\phi}_0 &=& (\omega_0-\mu) + s A_0 \rho_0^2 + 2 s B \rho_1^2
+ s B \rho_1^2 \cos(2 \varphi) 
\nonumber
\\
&+& s \left(\frac{\Gamma_1 \rho_1^3}{\rho_0}
+ 3 \rho_0 \rho_1 \Gamma_0 \right) \cos(\varphi), 
\label{beq7}
\end{eqnarray}
while the equations for $\rho_1, \phi_1$ are found by
interchanging indices $1$ and $0$ in the above equations. 
Next, taking into regard the conservation of the total norm, we obtain the equation 
$\rho_0^2+ \rho_1^2=N$, where $N=\int |u|^2 dx$ is the integral of motion of Eq. (\ref{beq1})
(the number of particles in BECs, or the power in optics). Finally, subtracting
Eq. (\ref{beq7}) for $\dot{\phi}_0$, and the corresponding one for $\dot{\phi}_1$, we obtain:
\begin{eqnarray}
\dot{\varphi} &=& -\Delta \omega + s (A_0 \rho_0^2 -  A_1 \rho_1^2) 
\nonumber 
\\
&-& s B (2 + \cos(2 \varphi)) (\rho_0^2-\rho_1^2)-s \frac{\cos(\varphi)}{\rho_0 \rho_1}
\nonumber
\\
&\times&  \left[\Gamma_0 \rho_0^2
(\rho_0^2-3 \rho_1^2) + \Gamma_1 \rho_1^2 (3 \rho_0^2 - \rho_1^2)\right].
\label{beq9}
\end{eqnarray}
Equations (\ref{beq6}), (\ref{beq9}) is a dynamical system, which, in principle, 
can be thoroughly 
investigated using phase-space analysis (such an approach has 
been presented in \cite{smerzi,kiv2} for similar systems that were derived using 
different expansion of the field $u$). Here, we will focus on the 
fixed points of 
the system [corresponding to the nonlinear 
eigenstates of Eq. (\ref{beq1})], and analyze 
their stability and bifurcations. 

{\it Results.} 
Below we will analyze all possible cases ($s=\pm 1$, $x_0 = 0$, $x_0 \ne 0$)
for the double well of Eq. (\ref{beq2}) with $V_0=1$, 
$\Omega=0.1$, and $w=0.5$ (the results do not change qualitatively using different values). 

First we consider the case of attractive nonlinearity,  
i.e., $s=-1$, and a symmetric double well potential with $x_{0}=0$ 
(implying that $\Gamma_0=\Gamma_1=0$). In this case, 
the parameters involved in Eqs. (\ref{beq6}) and (\ref{beq9}) are found to be 
$A_0 = 0.09078$, $A_1=0.09502$, $B=0.08964$, $\omega_0=0.13282$ and $\omega_1=0.15571$. 
Then, it is readily observed that the possible real solutions of Eq. (\ref{beq6}) 
are $\rho_0=0$ and $\rho_1=0$, as well as $\varphi=0 \hspace{1.5mm} ({\rm mod} \hspace{1.5mm} \pi)$. 
The former two are continuations of the linear solutions in the nonlinear regime. 
However, the latter one is a non-trivial combination of
the two modes for $\varphi=\pi$ that results in an {\it asymmetric} pair of mirror-symmetric 
solutions \cite{jackson,zhigang}, emerging through a {\it pitchfork} bifurcation. 
From Eq. (\ref{beq9}), we obtain that this new branch of solutions
bifurcates from the symmetric branch $(\rho_0,\rho_1)=(\sqrt{N},0)$ for 
\begin{eqnarray}
N>N_c=\frac{\Delta \omega}{3 B-A_0},
\label{beq10}
\end{eqnarray}
and for $\mu<\mu_c=\omega_0-A_0 N=0.12115$.
These analytical predictions are in {\it excellent} agreement with 
the numerical results $\mu_c = 0.122 (\pm 0.001)$. It is also easy to see that the
anti-symmetric branch $(\rho_0,\rho_1)=(0,\sqrt{N})$ does not give
rise to such a bifurcation. The different branches of the full
numerical solutions (including the bifurcating ones) have been 
obtained through numerical fixed point algorithms solving the steady
state version of Eq. (\ref{beq1}), and using continuation of the
solutions over the parameter $\mu$. The results are shown in the top left panel of Fig.1, where 
the norm of the solutions $N=\int |u|^2 dx$ is shown as a function of the chemical potential 
(or propagation constant in optics) $\mu$. In addition, as expected from the nature of the bifurcation, 
the linear stability analysis has been used to illustrate the following: The emerging new asymmetric 
(i.e., ``symmetry breaking'') branch of solutions is stable, while the original symmetric branch 
is unstable beyond the bifurcation point due to a real eigenvalue $\lambda_r$ (see bottom left panel of Fig. 1). 

\begin{figure}[t]
\includegraphics[width=4.2cm,height=5.2cm]{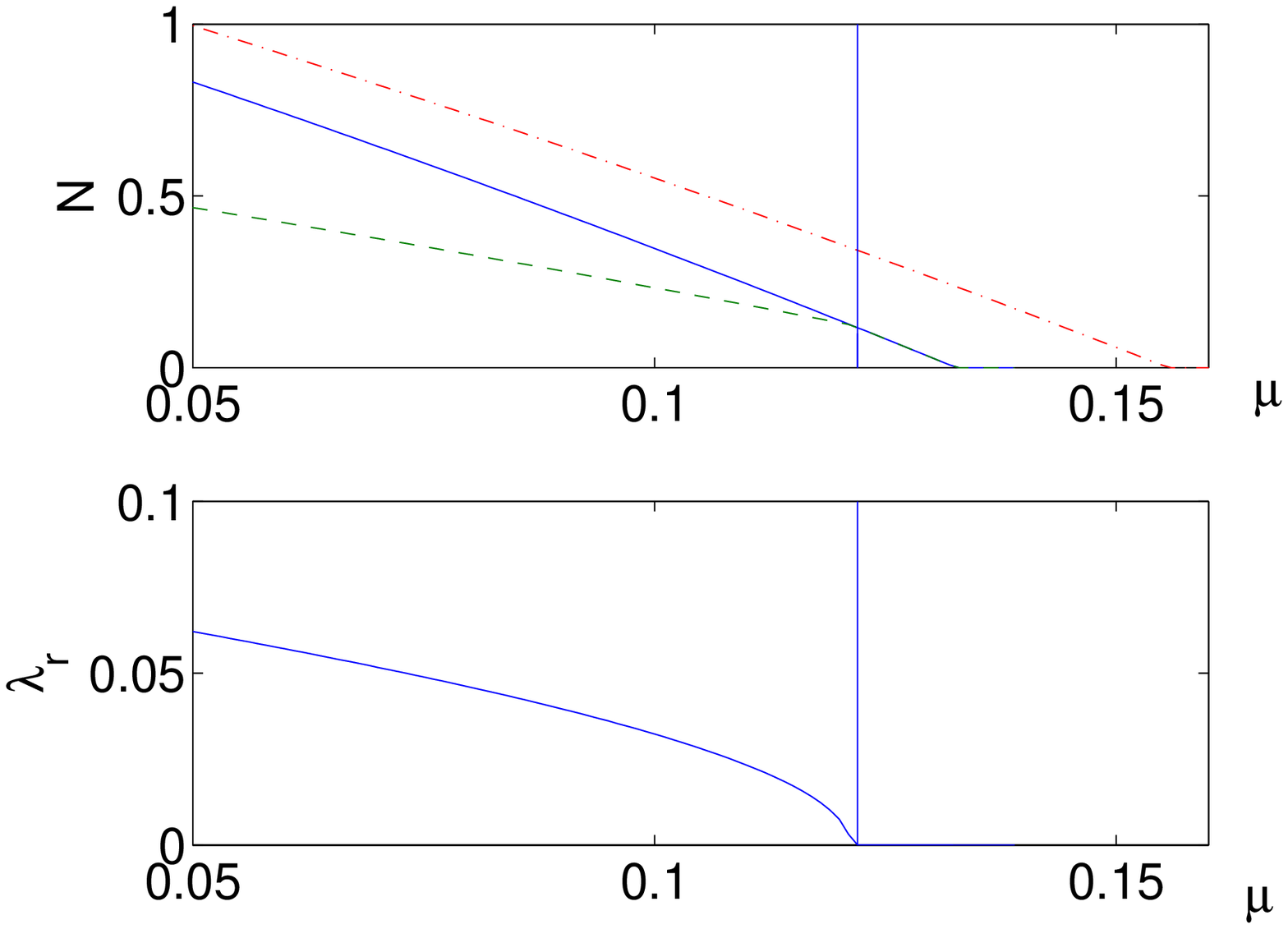}
\includegraphics[width=4.2cm,height=5.2cm]{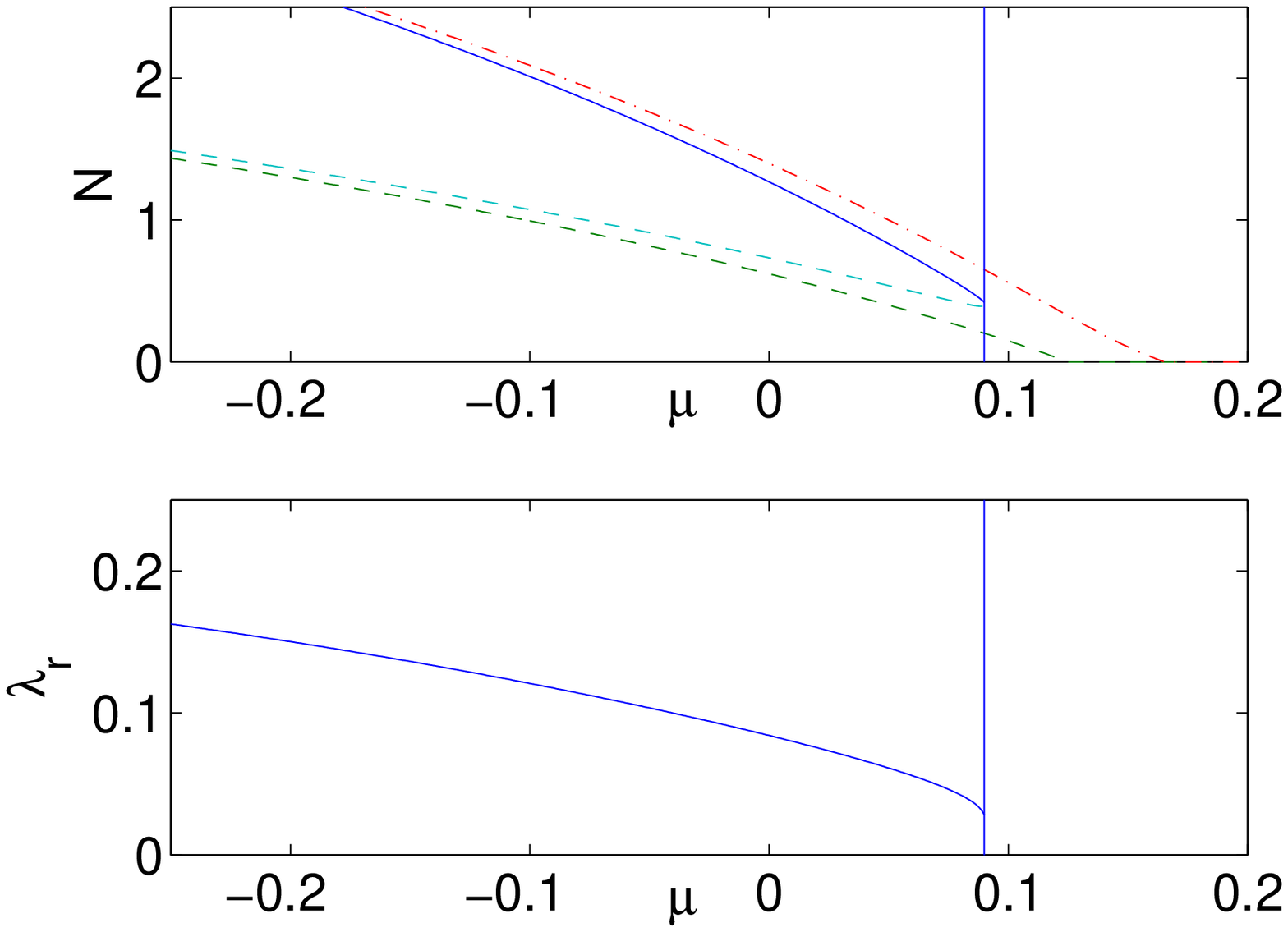}
\caption{(Color online) The top panels show the norm of the solutions of Eq.(\ref{beq1}) 
for attractive nonlinearity ($s=-1$) as a function of $\mu$ for symmetric 
(left panel, $x_{0}=0$) and asymmetric double well (right panel, $x_{0}=0.5$). 
The potential parameters are $\Omega=0.1$, $V_0=1$ and $w=0.5$. 
The 
solid lines denote the symmetric solution, the 
dashed-dotted lines denote the antisymmetric one, while the 
dashed lines 
denote the asymmetric solutions that are generated from the bifurcation 
at $\mu_c \approx 0.122$ (pitchfork) and $\mu_c \approx 0.009$ (saddle-node) respectively. 
The bottom panels show the maximal real eigenvalue associated with the linear stability of the symmetric branches.
}
\label{fig1}
\end{figure}

Next, in the same case ($s=-1$), we consider 
a double well potenial with a {\it weak asymmetry} ($x_{0}=0.5$). 
In this case, the constants involved in Eqs. (\ref{beq6})-(\ref{beq9}) are found to be 
$A_0 = 0.14903$, $A_1=0.15618$, $B=0.02958$, $\omega_0=0.1249$ and $\omega_1=0.16535$, while 
$\Gamma_0=0.0407$ and $\Gamma_1=-0.04077$. Note that even such a weakly 
asymmetric case,  
renders the right well ``shallower'', in the following sense: the density, or power $N$ 
(regarding the ground state of the linear problem), in the right well is smaller than the one in the left well. 
Thus, in the nonlinear problem, the respective branches that bear the larger part 
of the density in the right or in the left well (i.e., the ones having, roughly speaking, 
the shape of a single pulse in each of the wells) are no longer equivalent. 
This results in a {\it significant} difference between the asymmetric and the symmetric case discussed above, 
namely there is no longer a pitchfork bifurcation, but instead, there is a {\it saddle-node} bifurcation. 
This result is shown in the top right panel of Fig. 1, where $N$ is shown as a function of $\mu$. 
It is readily observed that, due to the nature of the saddle-node bifurcation, two branches 
(one of which is stable and the other one is unstable) 
``collide'' at some critical value of $\mu=\mu_c$ (see below) and disappear. These branches are 
the more ``symmetric'' one, that has support in both wells (see solid line in top left panel of Fig. 2), and the one 
pertaining to the state having the form of a single pulse in the shallower well (see dashed line in the 
rightmost top panel of Fig. 2). The instability of the former branch is depicted in the bottom right panel of Fig. 1, 
where the maximal real eigenvalue is shown as a function of $\mu$. 
On the other hand, there exists also another single-pulse branch 
supported over the deeper well (see top third panel of Fig. 2), which persists all the way to the linear limit. 
Furthermore, the dash-dotted anti-symmetric branch of the top right of Fig. 1 is shown
in the second top panel of Fig. 2. 

The novel feature described above, namely the asymmetric breakdown of the pitchfork bifurcation into 
a saddle-node one, 
is a {\it particular feature} of asymmetric double well potentials 
that, to the best of our knowledge, has not been previously appreciated.
Notice that Eq. (\ref{beq9}) 
predicts that the saddle-node bifurcation occurs at $\mu_c=0.08748$, 
while the numerical result is $\mu_c=0.09 \pm 0.001$; apparently the two results are again in excellent agreement. 

\begin{figure}[t]
\includegraphics[width=4.22cm,height=5.2cm]{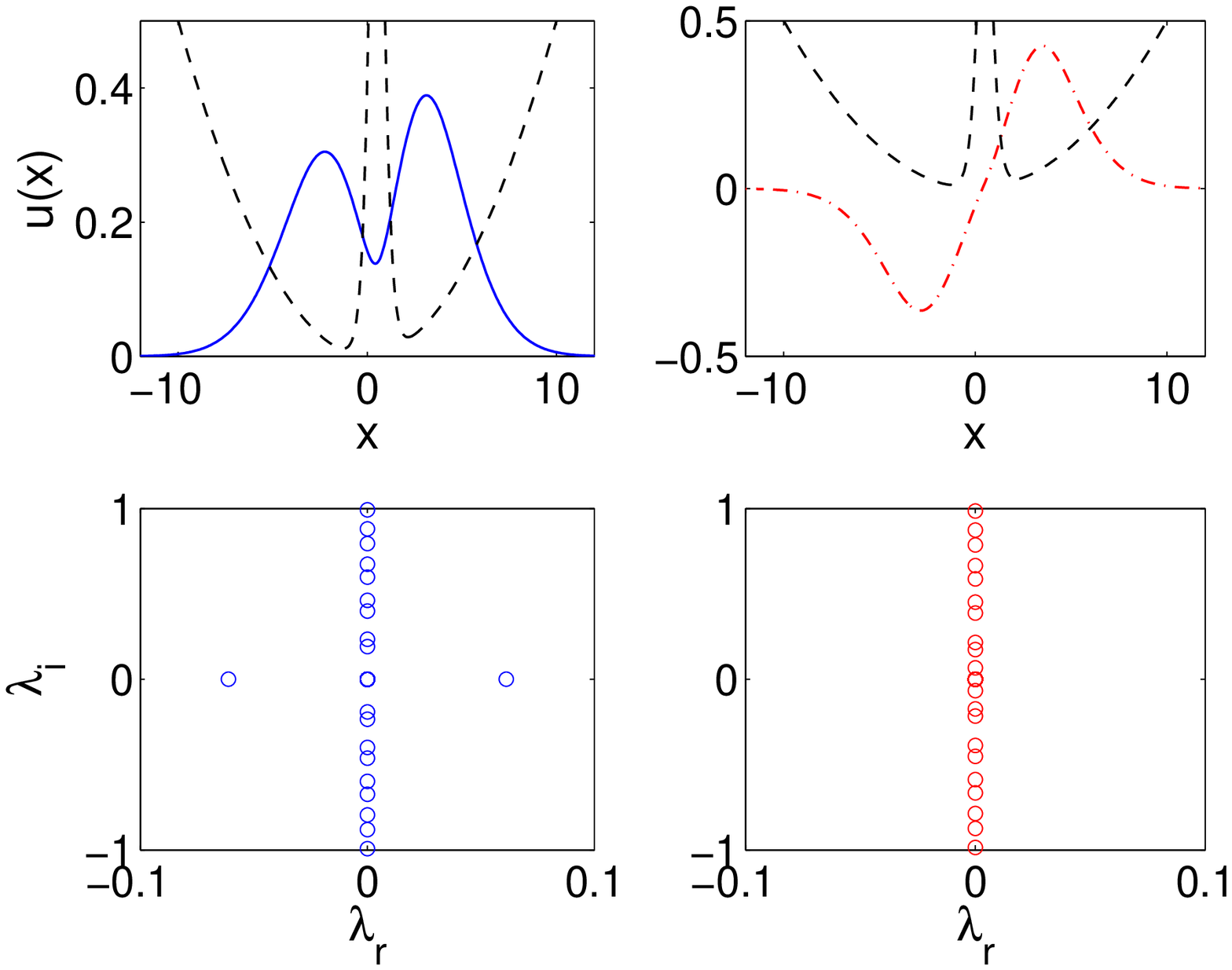}
\includegraphics[width=4.22cm,height=5.1cm]{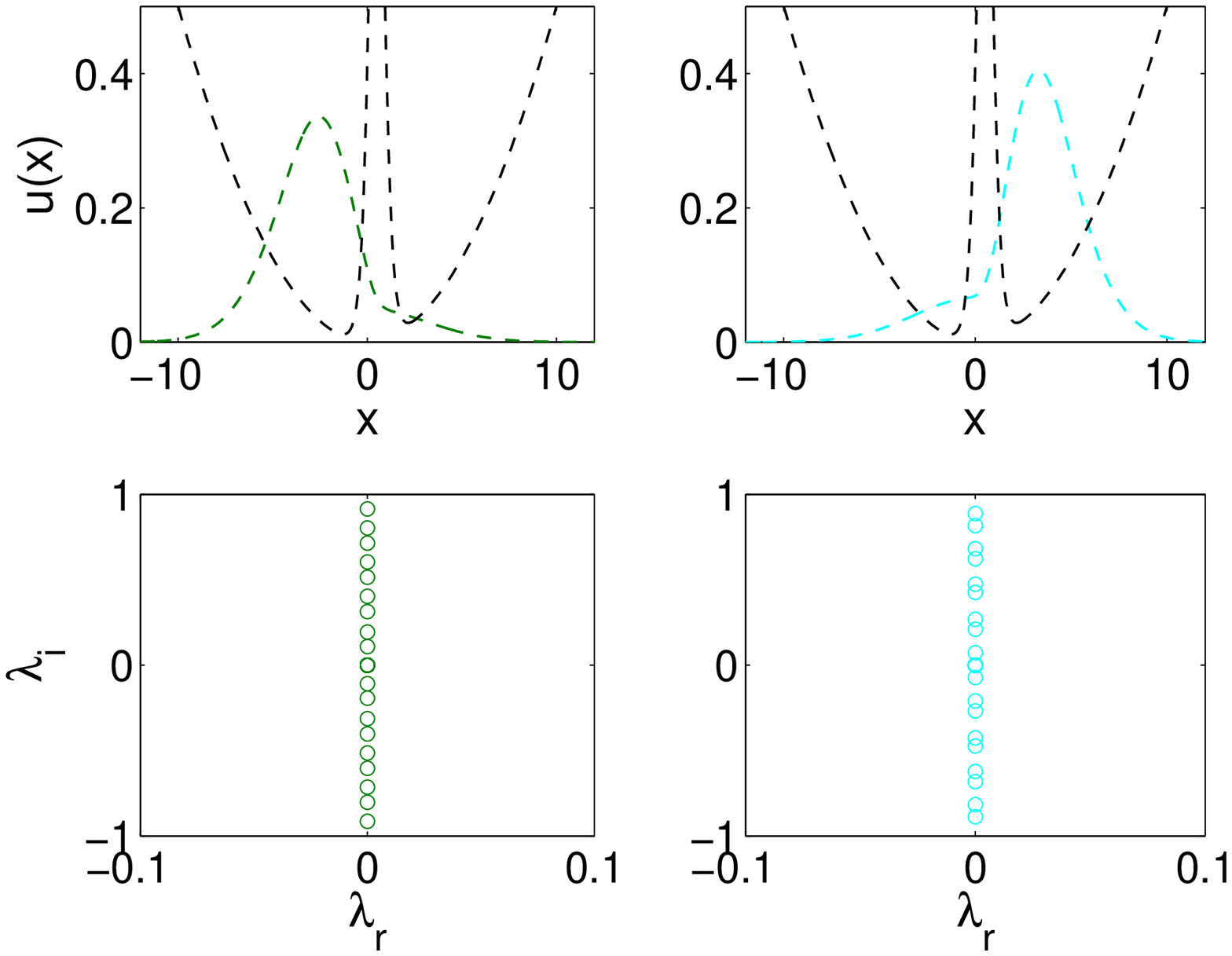}
\caption{(Color online) The steady state solutions of Eq.(\ref{beq1}), (see also Fig. \ref{fig1}) 
for the focusing, asymmetric case (top panels) and their linear stability (bottom panels) for $\mu=0.05$. 
The black-dashed line shows the double well potential.}
\label{fig2b}
\end{figure}

\begin{figure}[t]
\includegraphics[width=4.25cm,height=5.2cm]{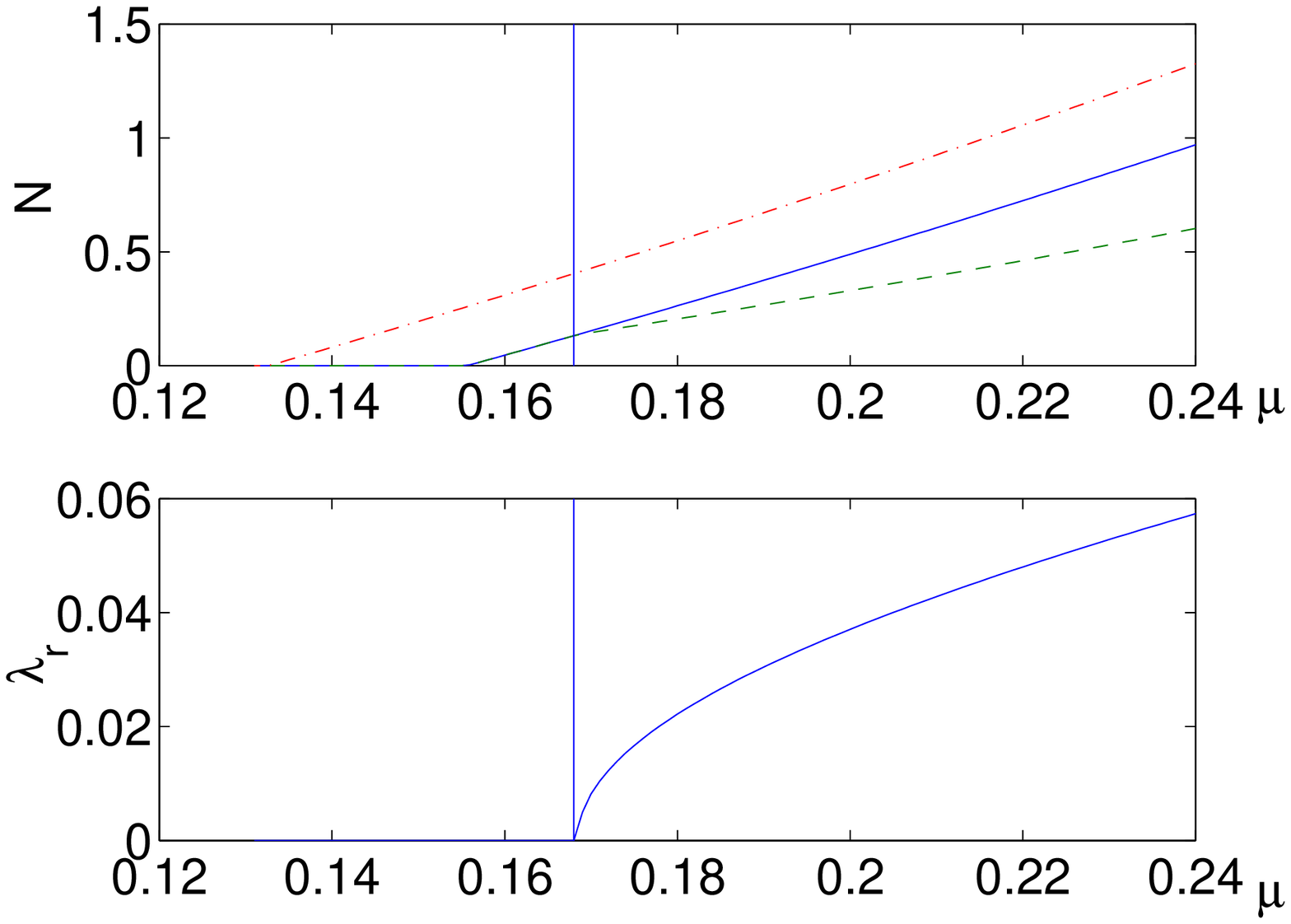}
\includegraphics[width=4.25cm,height=5.2cm]{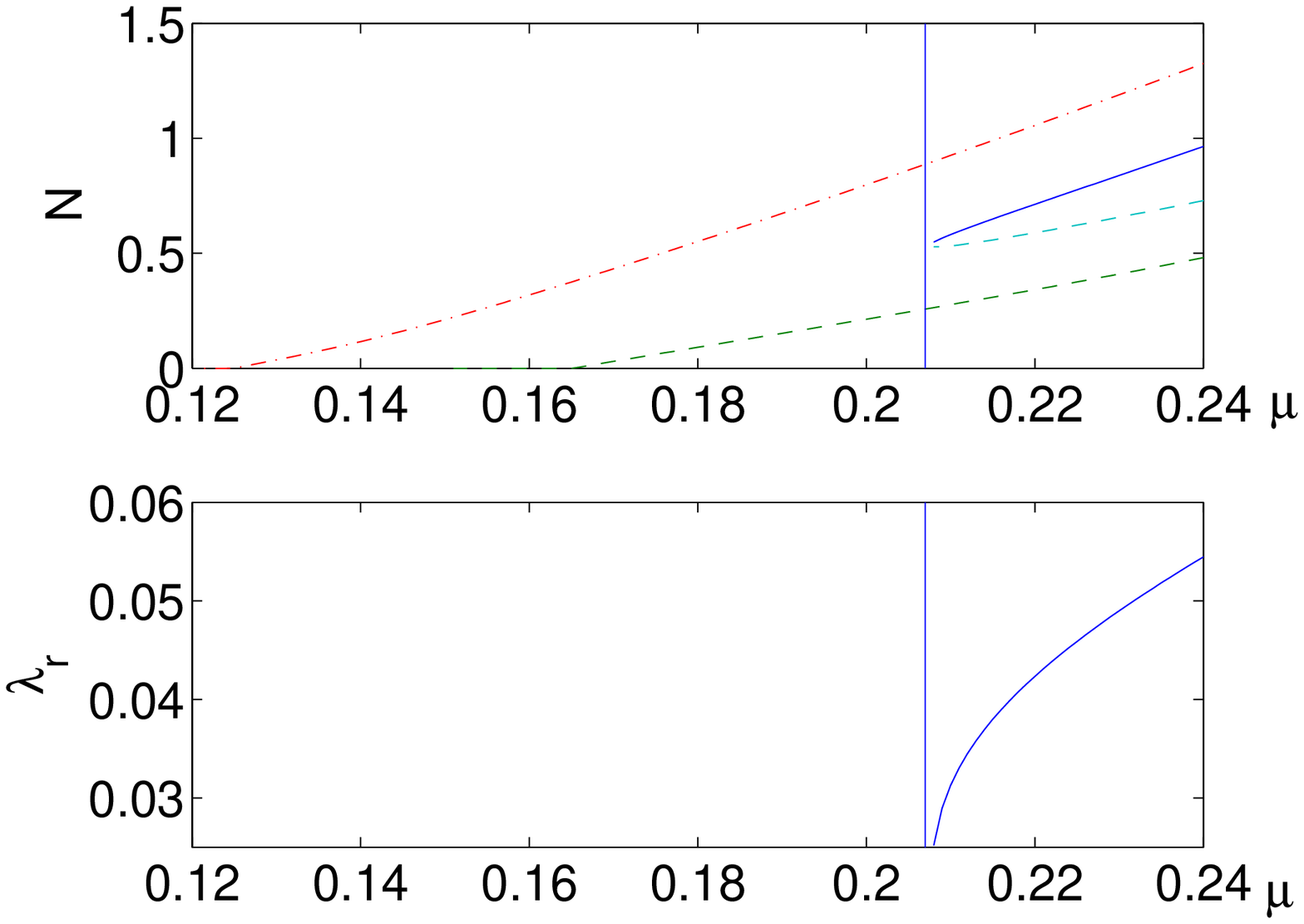}
\caption{(Color online) 
Same as in Fig.\ref{fig1} but for the repulsive nonlinearity ($s=+1$). 
The 
dashed-dotted lines denote the symmetric solution, the 
solid the antisymmetric, while the 
dashed lines
denote the asymmetric solutions generated from the bifurcation at 
$\mu_c \approx 0.168$ (pitchfork) and $\mu_c \approx 0.207$ (saddle-node) 
respectively. Contrary to the case $s=-1$, the bifurcations originate from the anti-symmetric branch.}
\label{fig3}
\end{figure}

Let us now consider the repulsive nonlinearity ($s=+1$). 
In this case, for the symmetric potential ($x_{0}=0$), the pitchfork bifurcation still occurs; 
however, now it does not originate from the symmetric branch, but rather from the anti-symmetric one with 
$(\rho_0,\rho_1)=(0,\sqrt{N})$, giving again rise to symmetry breaking. 
Analyzing Eq. (\ref{beq9}), we find that this occurs when
\begin{eqnarray}
N> N_c \geq \frac{\Delta \omega}{3 B- A_1}, 
\label{beq11}
\end{eqnarray}
and for $\mu=\omega_0 + 3 B N=0.16822$, once again in excellent agreement
with the numerical result $\mu_c= 0.168 (\pm 0.001)$.

\begin{figure}[t]
\includegraphics[width=4.25cm,height=5.2cm]{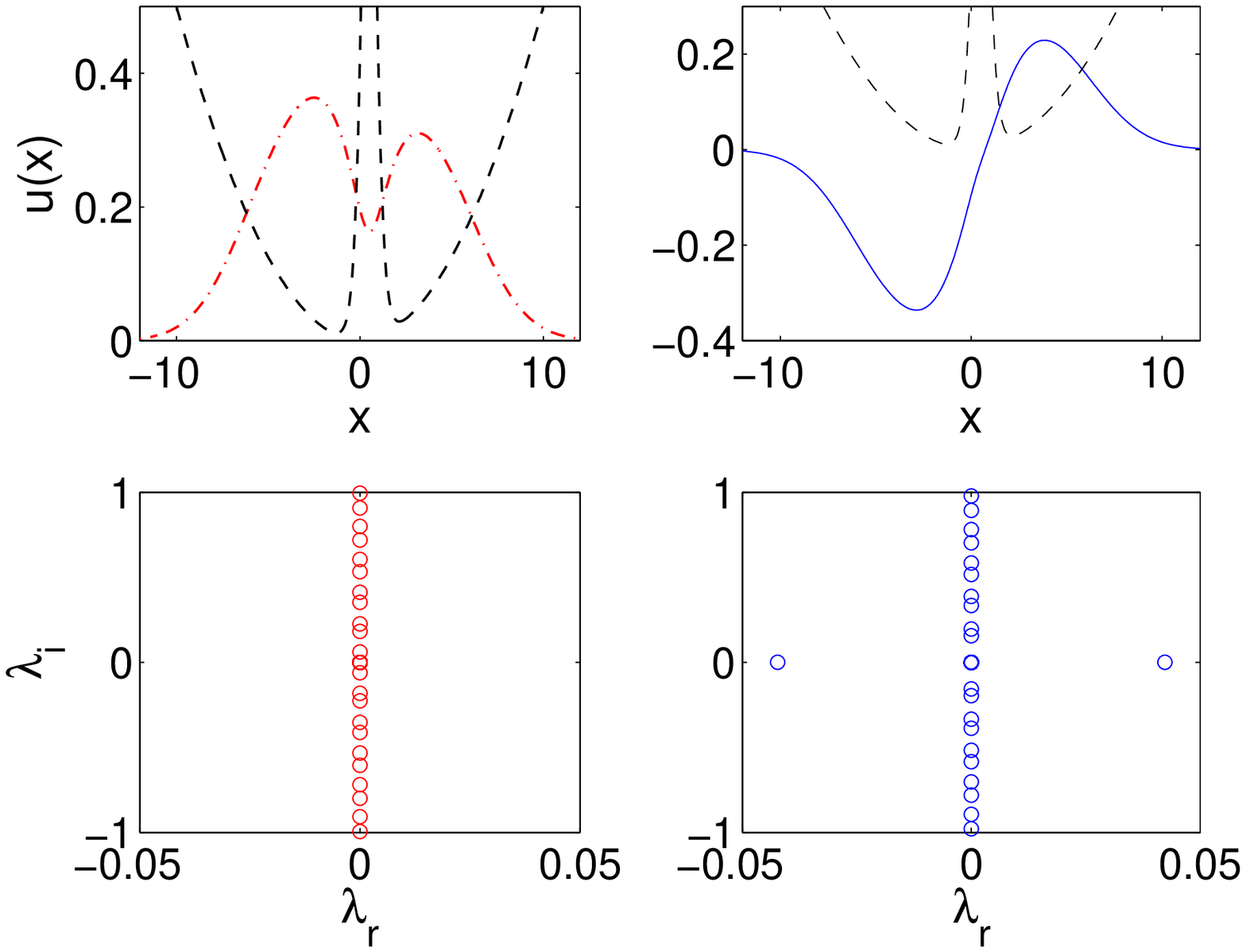}
\includegraphics[width=4.25cm,height=5.2cm]{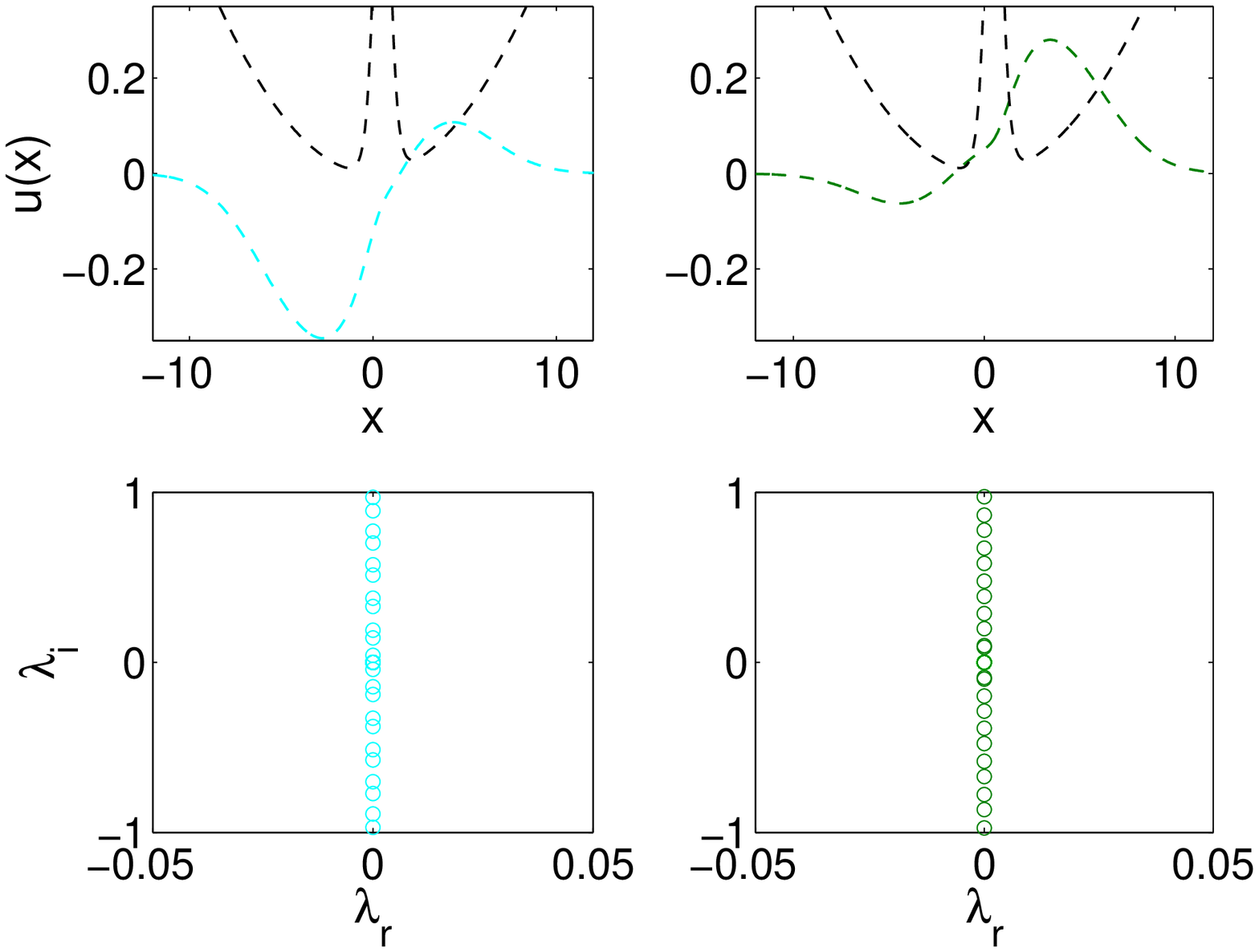}
\caption{(Color online) The steady state solutions of Eq.(\ref{beq1}), 
(see also Fig. \ref{fig3}) for $s=+1$ in the asymmetric case (top panels) and their linear 
stability (bottom panels) for $\mu=0.22$. The black-dashed line shows the double well potential.}
\label{fig4b}
\end{figure}

On the other hand, in the same case ($s=+1$) but for an asymmetric ($x_{0}=0.5$) 
double well, the bifurcation still originates from anti-phase (between the wells)
solutions, but again (as in the asymmetric case with $s=-1$), the bifurcation 
is of the saddle-node type. 
This  is theoretically predicted to occur at $\mu_c=0.21342$, 
once again in very close
agreement to the numerical result $\mu_c=0.207 \pm 0.001$. 
The details of the bifurcation diagrams, are illustrated in Fig. \ref{fig3} (analogously to Fig. 1), 
while the steady state solutions and their linear stability are shown in Fig. \ref{fig4b} (analogously to Fig. 2).


{\it Conclusions.} In conclusion, we have presented a systematic
analysis based on a Galerkin, two-mode truncation of the 
stationary states of symmetric and asymmetric double well potentials.
The analysis has been carried out both for repulsive and attractive nonlinearities and, 
as such, can be relevant to a variety of physical contexts; these include matter-wave physics 
(most directly), nonlinear optics, as well as other contexts where 
it is relevant to consider double well potentials in the NLS model proper.
We have demonstrated that our analytical approach describes quite accurately, both 
{\it qualitatively} and {\it quantitatively} the features of the nonlinear solutions; 
numerical results were shown to be in excellent agreement with the analytical 
predictions. 

In the case of a symmetric double well potential, it has been shown 
that a symmetry-breaking (pitchfork) bifurcation of the ground state occurs 
for attractive nonlinearities, while it is absent for repulsive nonlinearities.
It has also been found that a similar bifurcation of the first excited state occurs 
in the relevant branches for repulsive nonlinearities, oppositely to the case of attractive ones,
where such bifurcation does not happen. 
Additionally, regarding the above feature, we have illustrated that 
symmetric potentials are very particular (degenerate) due to their 
special characteristic of mirror-equivalence of the emerging symmetry-breaking states.
We have shown that even weak asymmetries lift 
this degeneracy and lead to saddle-node bifurcations 
instead of pitchfork ones that were similarly quantified 
in both attractive and repulsive nonlinearity contexts. 

These results underscore the relevance of analyzing steady state features 
of nonlinear models (in the presence of external potentials) based 
on the states of the underlying linear equations. It would be 
particularly interesting to examine the extent to which dynamical 
features of such models can be captured by similar truncations. 
Such studies are currently in progress. 

{\it Acknowledgements.} Constructive discussions with M.K. Oberthaler are kindly acknowledged. 
This work has been partially supported from ``A.S. Onasis'' Public Benefit Foundation (G.T.) and NSF (P.G.K.).


\end{document}